\documentclass[final]{aipproc}

\layoutstyle{6x9}

 \newcommand\la{\langle}
 \newcommand\ra{\rangle}
 \newcommand\beq{\begin{equation}}
 
 \newcommand\eeq{\end{equation}}
 \newcommand\beqn{\begin{eqnarray}}
 \newcommand\eeqn{\end{eqnarray}}
 \newcommand\GeV{{\rm GeV}}
 
\def\GeV{\,\mbox{GeV}}

\def\lsim{\mathrel{\rlap{\lower4pt\hbox{\hskip1pt$\sim$}}
    \raise1pt\hbox{$<$}}}         
\def\gsim{\mathrel{\rlap{\lower4pt\hbox{\hskip1pt$\sim$}}
    \raise1pt\hbox{$>$}}}         
\def\BA{\begin{eqnarray}}
\def\BE{\begin{equation}}
\def\BF{\begin{figure}[htb]}
\def\BT{\begin{table}[htb]}
\def\EA{\end{eqnarray}}
\def\EE{\end{equation}}
\def\EF{\end{figure}}
\def\ET{\end{table}}

\def\la{\langle}
\def\ra{\rangle}

\begin{document}

\title{
Forward Physics in Proton-Nucleus and Nucleus-Nucleus Collisions}



\author{J.~Nemchik}{
address={Institute of Experimental Physics SAS, Watsonova 47,
04001 Kosice, Slovakia},
altaddress={
Czech Technical University,
FNSPE, Brehova 7,
11519 Praque, Czech Republic},
}

\author{I.K.~Potashnikova}{
address={
Departamento de F\'{\i}sica y Centro de Estudios Subat\'omicos,\\ 
Universidad T\'ecnica Federico Santa Mar\'{\i}a, Casilla 110-V,
Valpara\'{\i}so, Chile}
}

\copyrightyear  {2008}

\begin{abstract}
We present an universal treatment for 
a substantial nuclear suppression representing
a common feature
of all known reactions on nuclear targets (forward production of 
high-$p_T$ hadrons, production of direct photons, the Drell-Yan
process, heavy flavor production, etc.).
Such a suppression at large Feynman $x_F$, corresponding to region
of minimal light-cone momentum fraction variable $x_2$ in nuclei,
is tempting to interpret as a manifestation of coherence
or the Color Glass Condensate. 
We demonstrate, however, that it is actually a simple consequence
of energy conservation and takes place even at low energies,
where no effects of coherence are possible.
We analyze this common suppression mechanism for several
processes performing
model predictions in the light-cone
dipole approach.
Our calculations agree with the data.

\end{abstract}

\date{July 8st, 2008}
\maketitle


%
%
\section{Introduction}
\label{intro}
%
%

In the proton(deuteron)-nucleus and nucleus-nucleus collisions,
investigated at the Relativistic Heavy Ion
Collider (RHIC), 
recent measurements of high-$p_T$ particle spectra
by the BRAHMS \cite{brahms,brahms-07},
STAR \cite{star} and PHENIX \cite{phenix}
Collaborations show a strong nuclear suppression. 
Observed nuclear effects occur not only
at large forward rapidities \cite{brahms,brahms-07,star}
but unexpectedly also at midrapidities \cite{phenix}.

Besides, quite strong and universal nuclear suppression
at large Feynman $x_F$
is confirmed by the collection of data from
\cite{soft} for the production of different
species of particles in $p-A$ collisions.
The rise of the nuclear suppression with $x_F$
is also supported by the NA49 data \cite{na49}
at lower energy corresponding to c.m.s.
energy $\sqrt{s} = 17.3\,$GeV.
The onset of strong nuclear effects at large $p_T$
has been also demonstrated for direct photon production
in $Au-Au$ collisions at RHIC by the
PHENIX Collaboration \cite{phenix-dp}.
The E772 experiment at Fermilab \cite{e772}
first observed that the Drell-Yan (DY) process
is considerably suppressed at large $x_F$. 

Assuming large forward rapidities,
the basic explanation for such an effect has been based
on an idea that in this kinematic region
corresponding to the beam fragmentation region
at large Feynman $x_F$ one can reach the smallest values
of the light-front momentum fraction variable $x_2$ in nuclei.
It allows to access the strongest coherence effects
such as those associated with shadowing or the Color Glass
Condensate (CGC). 

It was shown in refs.~\cite{knpsj-05,knpsj-05c,npps-08} 
that a considerable nuclear
suppression for any reaction at large $x_F$ (small $x_2$)
is caused by another effects, which can be easily
misinterpreted as coherence. 
Such a suppression 
can be treated, alternatively,
as a Sudakov suppression, a consequence of a reduced
survival probability for large rapidity gap (LRG) processes
in nuclei, an enhanced resolution of higher Fock
states by nuclei, or an effective energy loss that
rises linearly with energy.
It was demonstrated in refs.~\cite{knpsj-05,knpsj-05c} that the nuclear
suppression at large $x_F$ is a leading
twist effect, violating QCD factorization.

In this paper we will analyze nuclear suppression at
large rapidities (large $x_F$)
for the following processes occurring 
in $p(d)-A$ and $A-A$ collisions: 
\begin{itemize}

\item
production of leading hadrons with small $p_T$

\item
high-$p_T$ hadron production at forward rapidities in p(d)-A
collisions 

\item
production of hadrons at small energies vs. NA49 data

\item
high-$p_T$ hadron production at midrapidities 

\item
direct photon production in Au-Au collisions 

\item
Drell-Yan production at large $x_F$
\vspace*{-0.1cm}
\end{itemize}

%
%
\section{Survival probability of large rapidity gaps}\label{sudakov}
%
%

Treating any hard reaction, which is 
LRG process in the limit
$x_F\to 1$, gluon radiation is forbidden by energy conservation.
If a large-$x_F$ particle is produced, the rapidity interval to be kept
empty is $\Delta y=-\ln(1-x_F)$. 
Assuming as usual an uncorrelated Poisson distribution for
gluons, the Sudakov suppression factor, i.e. the probability to have a
rapidity gap $\Delta y$, becomes
%
%
 \beq
S(\Delta y) = e^{-\la n_G(\Delta y)\ra}\ ,
\label{20}
 \eeq
%
%
where $n_G(\Delta y)$ is the mean number of gluons that would
be radiated within $\Delta y$ if energy conservation were not an issue.

The mean number $\la n_G(\Delta y)\ra$ of gluons radiated in the rapidity
interval $\Delta y$ is related to the height of the plateau in the gluon
spectrum, $\la n_G(\Delta y)\ra=\Delta y\,dn_G/dy$. Then, the Sudakov
factor acquires the simple form,
%
%
 \beq
S(x_F) = (1-x_F)^{dn_G/dy}\ .
\label{40}
 \eeq
%
%
The height of the gluon plateau was estimated in ref.~\cite{gb} as,
%
%
 \beq
\frac{dn_G}{dy} = \frac{3\alpha_s}{\pi}\,
\ln\left(\frac{m_\rho^2}{\Lambda_{QCD}^2}\right)\ .
\label{50}
 \eeq
%
%

For further calculations we take $\alpha_s=0.4$
(see discussion in ref.~\cite{knpsj-05}), which gives 
with high accuracy $dn_G/dy=1$, i.e. the Sudakov factor,
%
%
 \beq
S(x_F)=1-x_F\ .
\label{60}
 \eeq
%
%

One can formulate nuclear suppression as $x_F\to 1$ as a survival
probability of the LRG in multiple interactions with the nucleus.
Every additional inelastic interaction contributes an extra suppression
factor $S(x_F)$. The probability of an n-fold inelastic collision is
related to the Glauber model coefficients via the
Abramovsky-Gribov-Kancheli (AGK)  cutting rules \cite{agk}.
Then the survival probability at impact parameter $\vec b$
reads,
%
%
 \beqn
W^{hA}_{LRG}(b) &=&
\exp[-\sigma_{in}^{hN}\,T_A(b)]
\sum\limits_{n=1}^A\frac{1}{n!}\,
\left[\sigma_{in}^{hN}\,T_A(b)\,
\right]^n\,S(x_F)^{n-1}\ ,
 \label{70}
 \eeqn
%
%
where $T_A(b)$ is the nuclear thickness function.

%
%
\section{Production of leading hadrons with small $p_T$}
\label{soft}
%
%

The left panel of Fig.~\ref{small-pt} shows
the collection of data from \cite{soft}
for production of different species of particles in $p-A$
collisions exhibiting quite a strong and universal suppression
at large $x_F$. Moreover, these data cover the laboratory energy range from
70 to 400 GeV and demonstrate so the $x_F$ scaling of nuclear effects.

It is natural to relate the observed suppression to the dynamics
discussed in the previous section. The nuclear effects can be
calculated using Eq.~(\ref{70}) summing over the number of collisions
and integrating over the impact parameter,
%
%
 \beqn
R_{A/N}(x_F) &=&
\frac{1}{(1 - x_F)\,\sigma_{eff}\,A}\,
\int d\,^2\,b e^{- \sigma_{eff}\,T_A(b)}\,
\Bigl \{e^{(1 - x_F) \sigma_{eff} T_A(b)} - 1\Bigr \}.
 \label{80}
 \eeqn
%
%
In the Glauber model
$\sigma_{eff} = \sigma_{in}^{NN}$.
However, Gribov's inelastic shadowing corrections
substantially reduce $\sigma_{eff}$ \cite{mine,kps-06}.

To compare with data, the nuclear effects are parametrized
as $R_{A/N}\propto A^{\alpha}$, where the exponent $\alpha$
varies with $A$. We used $A = 40$, for which
the Gribov corrections evaluated in \cite{kps-06} lead
to $\sigma_{eff}\sim 20\,$mb. Then a simple expression
Eq.~(\ref{80}) explains the observed
$x_F$ scaling and describes rather well the data.

%
%
\section{High-$p_T$ hadron production at forward rapidities}
\label{scat}
%
%

 \begin{figure}[tbh]
\resizebox{13.3pc}{!}{
    \includegraphics[height=0.2\textheight]{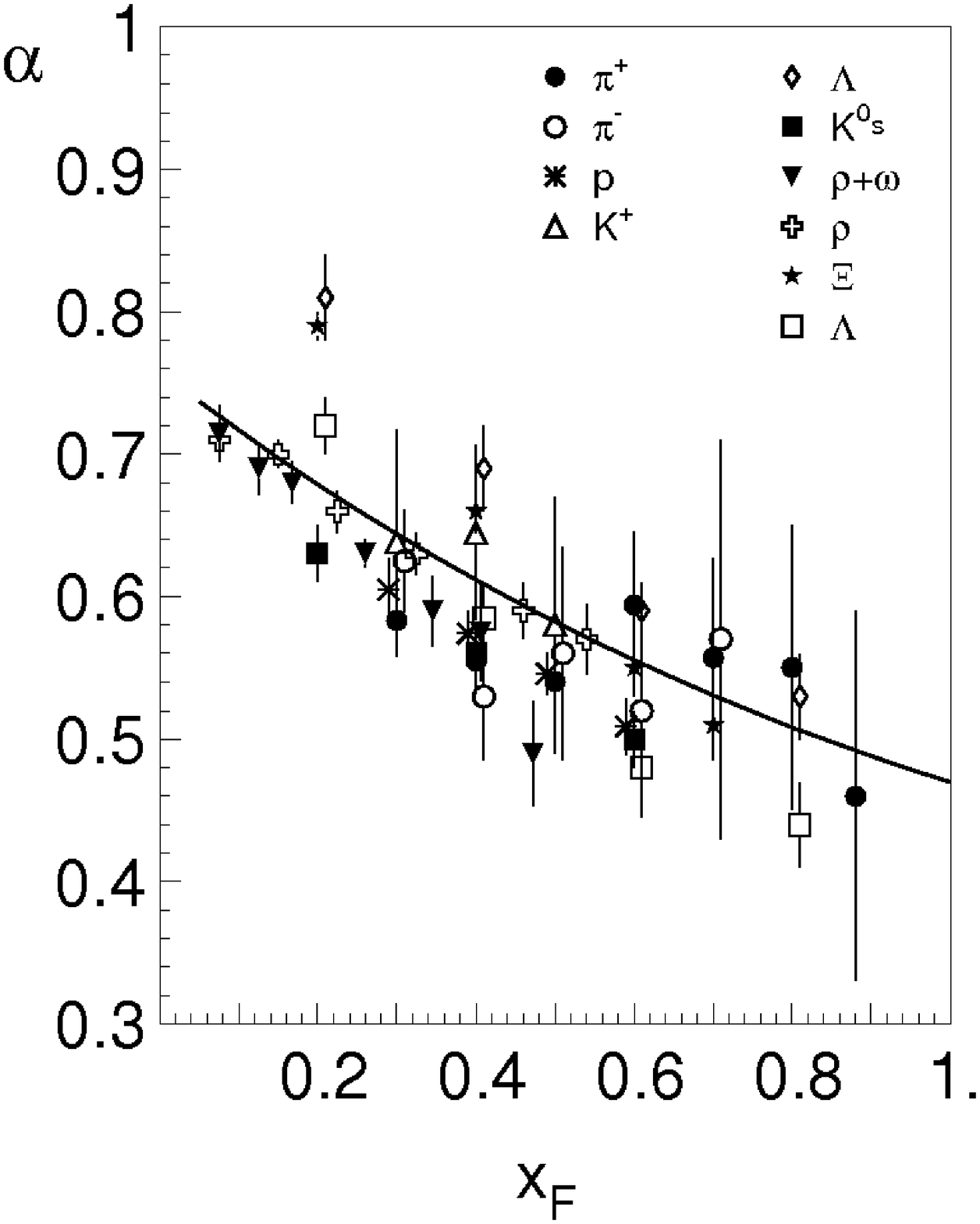}}
\hspace*{2.2cm}
\resizebox{15pc}{!}{
     \includegraphics[height=0.4\textheight]{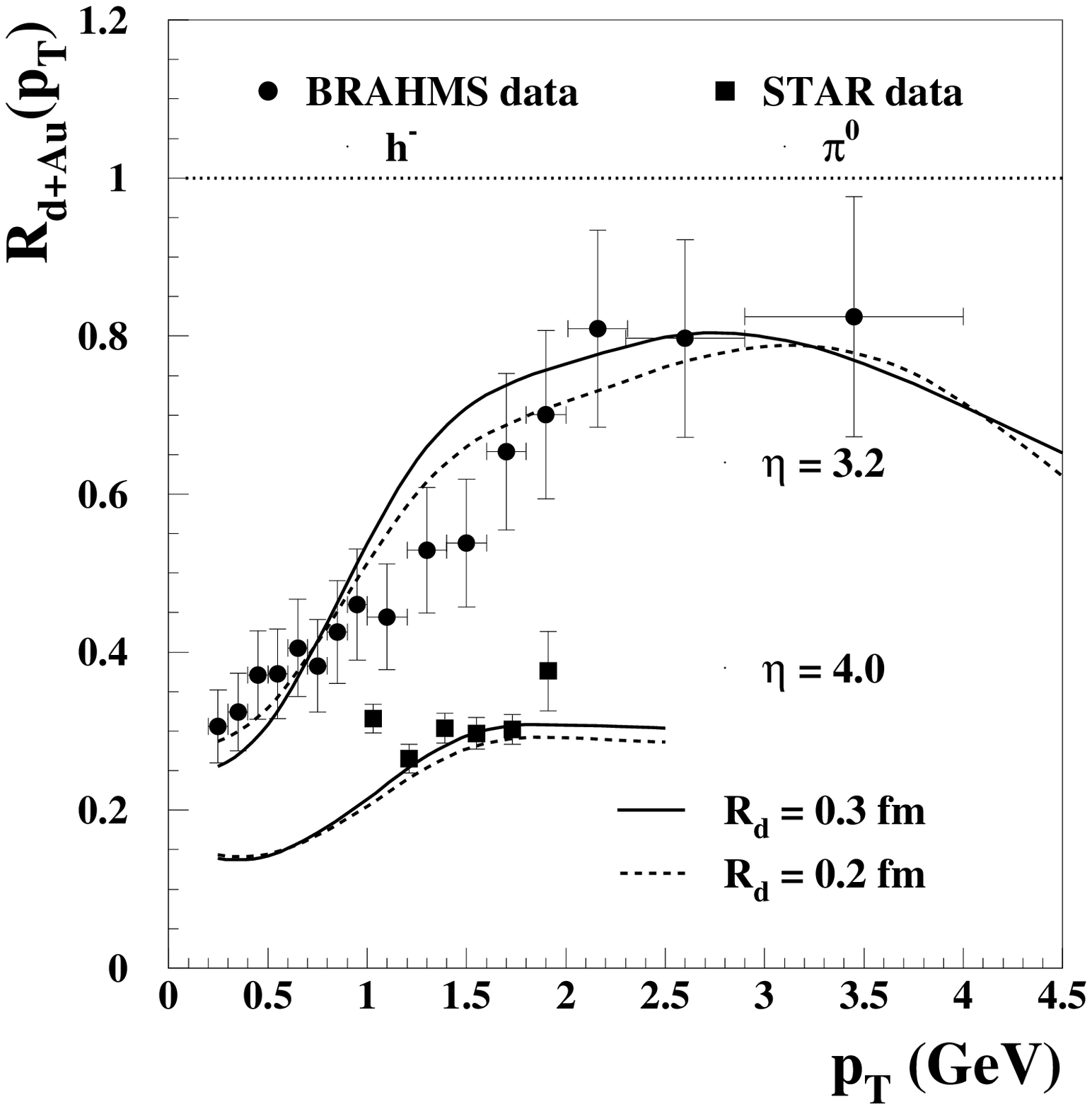}}
\caption
 {
(Left)
Exponent describing the $A$
dependence $(\propto A^\alpha)$
of the nucleus-to-proton ratio for production of different
hadrons as a function of $x_F$. 
(Right)
Ratio of negative hadron and neutral pion production rates 
in $d-Au$ and $pp$
collisions as function of $p_T$ at pseudorapidity $\eta = 3.2$ and
$\eta = 4.0$ vs. data 
from the BRAHMS \cite{brahms} and STAR Collaborations \cite{star},
respectively.
}
 \label{small-pt}
 \end{figure}

The cross section of hadron production in $dA(pp)$ collisions is given by
a convolution of the distribution function for the projectile valence
quark with the quark scattering cross section and the fragmentation
function,
%
%
 \beqn
&&\frac{d^2\sigma}{d^2p_T\,d\eta} =
\sum\limits_q \int\limits_{z_{min}}^1 dz\,
f_{q/d(p)}(x_1,q_T^2)
\left.\frac{d^2\sigma[qA(p)]}{d^2q_T\,d\eta}
\right|_{\vec q_T=\vec p_T/z}\,
D_{h/q}(z),
\label{85}
 \eeqn
%
%
where
$x_1=\frac{q_T}{\sqrt{s}}\,e^\eta$.
The quark distribution functions in the nucleon have the form using
the lowest order parametrization of Gluck, Reya and Vogt \cite{grv}. 
We used proper fragmentation functions
using parametrization from \cite{fs-07}.

Interaction with a nuclear
target does not obey factorization, since the effective projectile quark
distribution correlates with the target. 
The main source of suppression at large $p_T$ concerns
to multiple quark rescatterings
in nuclear matter.
Summed over multiple interactions, the quark distribution in the
nucleus reads,
%
%
 \beqn
\hspace*{-0.40cm}
f^{(A)}_{q/N}(x_1,q_T^2)&=&C\,f_{q/N}(x_1,q_T^2)\,
\frac{\int d^2b\,
\left[e^{-x_1\sigma_{eff}T_A(b)}-
e^{-\sigma_{eff}T_A(b)}\right]}
{(1-x_1)\int d^2b\,\left[1-
e^{-\sigma_{eff}T_A(b)}\right]}\, ,
\label{100}
 \eeqn
%
%
where the effective cross section
$\sigma_{eff} = \sigma_{eff}(p_T,s)=\left\la
\sigma^2_{\bar qq}(r_T)\right\ra /
\left\la
\sigma_{\bar qq}(r_T)\right\ra$
has been evaluated in \cite{knpsj-05}.
The normalization factor $C$ in Eq.~(\ref{100}) is fixed by the Gottfried
sum rule.

The cross section of quark scattering on the target
$d\sigma[qA(p)]/d^2q_Td\eta$
in Eq.~(\ref{80}) is
calculated in the light-cone dipole approach \cite{zkl,jkt-01}.
In our
calculations, we separate the contributions characterized by different
initial transverse momenta and sum over different mechanisms of
high-$p_T$ production. Details can be found in \cite{knpsj-05}.

The BRAHMS Collaborations \cite{brahms} in 2004
found a substantial nuclear suppression for high-$p_T$
negative hadrons produced at pseudorapidity $\eta = 3.2$.
Two years later, the STAR Collaboration \cite{star} has been
observed even stronger suppression for neutral pions 
at $\eta = 4.0$ as one see from the right panel of Fig.~\ref{small-pt}. 
Because the data cover rather small $x_2\sim 10^{-3}$, the
interpretation of such a suppression has been tempted to be
as a result of saturation \cite{glr,al} or the
CGC \cite{mv}, expected in some models \cite{kkt}.

Even if one supposes to interpret the observed suppression
at $\eta = 3.2$
in terms of CGC, such an interpretation should fail
at larger $\eta = 4.0$, where
the observed suppression is more than a factor of 2 larger.
The stronger onset of the quantum coherence effects at 
$\eta = 4.0$ can not
explain such a huge rise of nuclear suppression.

Much stronger nuclear effects
at $\eta = 4$
can be simply explained
by the energy conservation as a much smaller
survival probability of LRG 
at larger $\eta$-values
\cite{knpsj-05,npps-08}.

\begin{figure}[h]
\resizebox{15pc}{!}{
     \includegraphics[height=0.4\textheight]{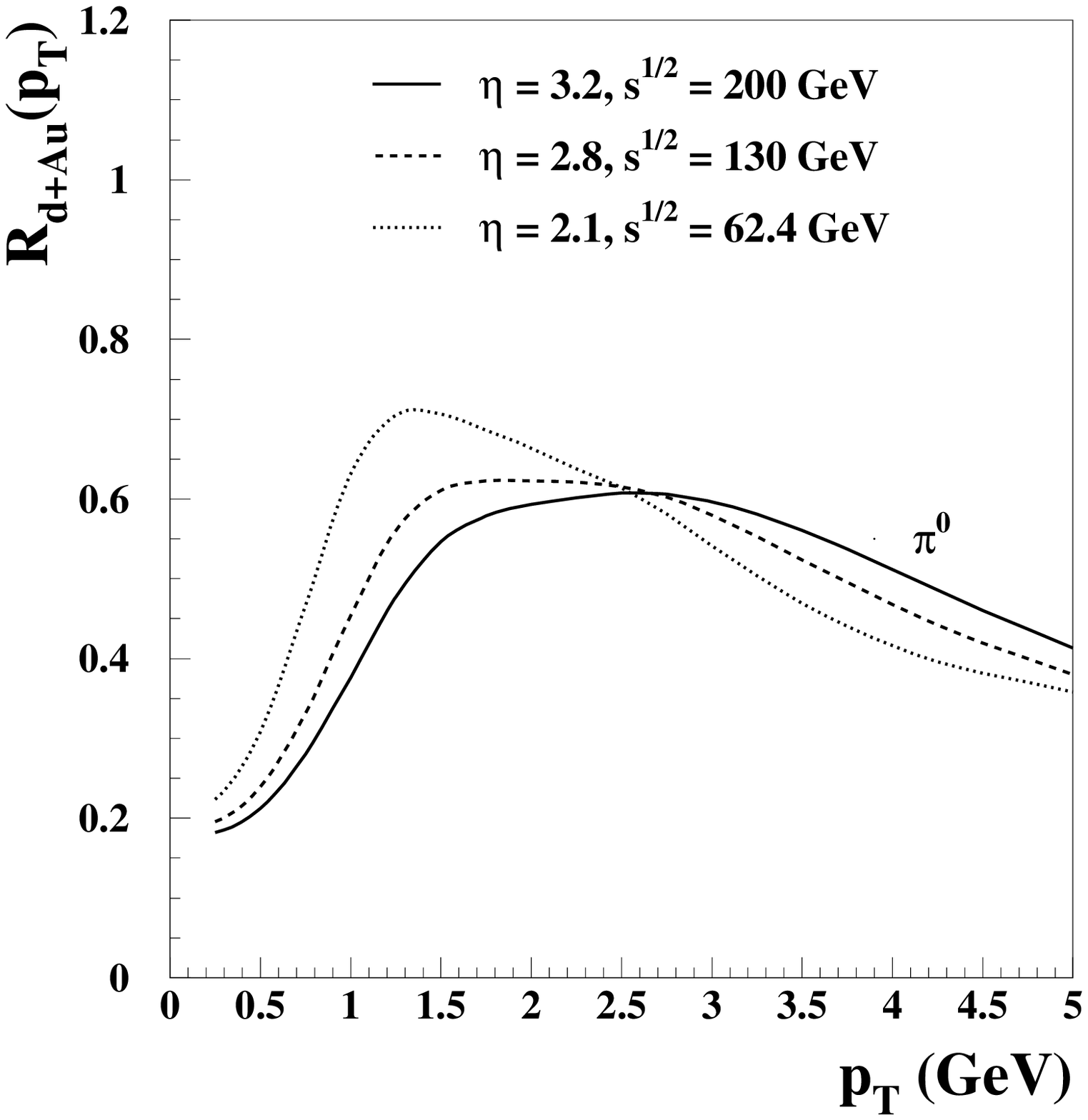}}
\hspace*{1.0cm}
\resizebox{15pc}{!}{
     \includegraphics[height=0.4\textheight]{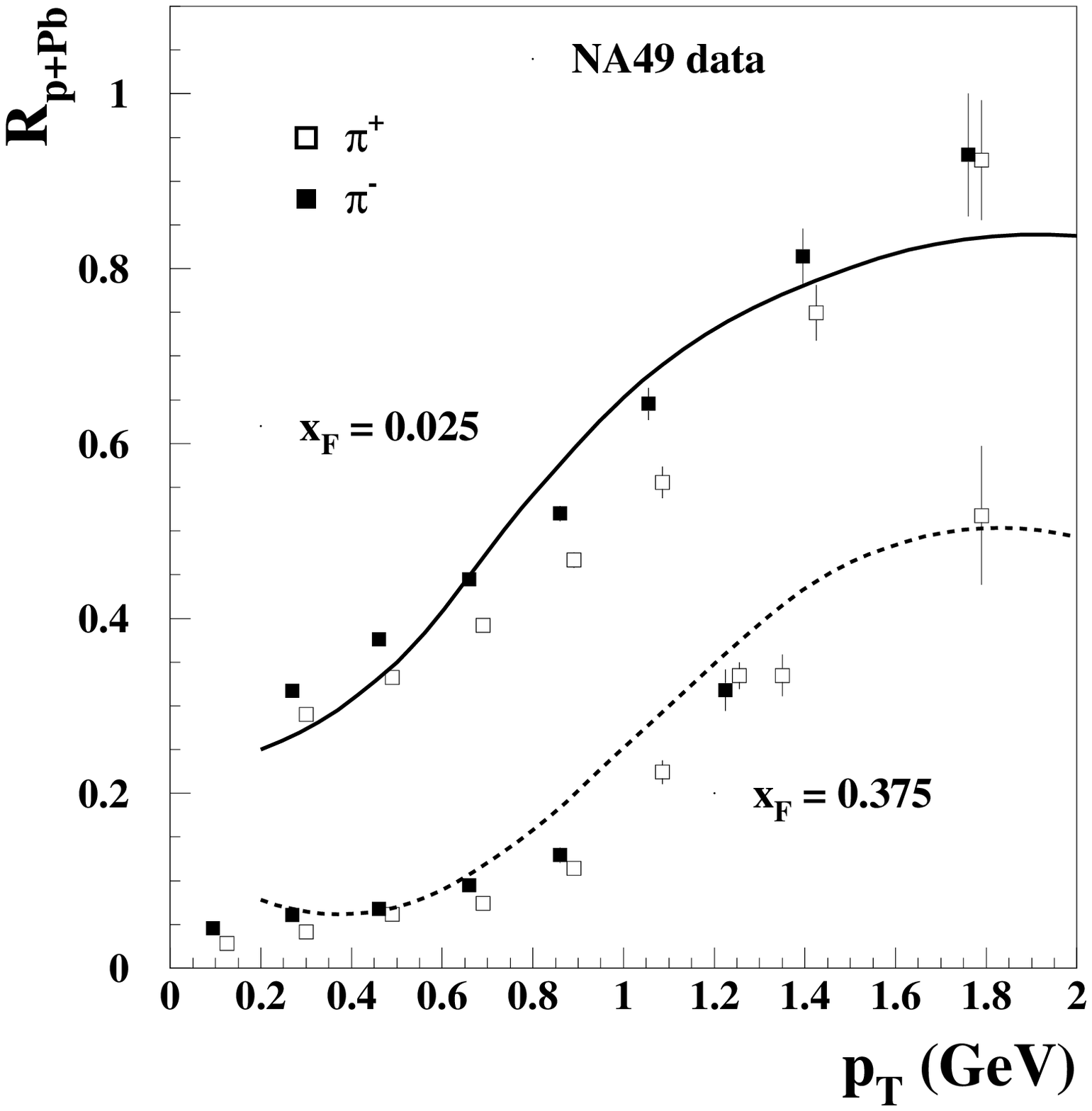}}
\caption
 {
(Left)
Theoretical predictions for an approximate $\exp(\eta)/\sqrt{s}$- scaling
of the ratio $R_{d+Au}(p_T)$ for $\pi^0$ production rates
in $d-Au$ and $pp$ collisions.
(Right)
Ratio, $R_{p+Pb}(p_T)$, for $\pi^\pm$ production rates
in $p-Pb$ and $pp$ collisions 
as function of $p_T$ at two fixed values of Feynman
$x_F = 0.025$ and $0.375$
vs. the NA49 data \cite{na49}.
}
 \label{brahms}
\end{figure}

Energy conservation applied for multiple parton
rescatterings leads to
$x_F$ scaling of nuclear effects
\cite{knpsj-05,knpsj-05c,npps-08}.
We expect approximately the same 
nuclear effects at different energies and pseudorapidities
corresponding to the same values of $x_F$. 
Such a situation is demonstrated in the left panel of Fig.~\ref{brahms},
where we present $p_T$ dependence of nuclear attenuation
factor $R_{d+Au}(p_T)$ for $\pi^0$ production
at different c.m.s. energies and $\eta$  
keeping the same value of $x_F$.
\vspace*{-0.2cm}

%
%
\section{Nuclear suppression at small energy vs. NA49 data}
\label{na49se}
%

The
right panel of Fig.~\ref{brahms} clearly demonstrates a stronger onset
of nuclear effects at larger $x_F$. 
The model predictions for nuclear suppression have been performed 
employing the dipole formalism and using the mechanisms
for the valence quarks described 
in~\cite{knpsj-05}.
One can see a good agreement of our calculations with
NA49 data \cite{na49}.
\vspace*{-0.2cm}

%
%
\section{High-$p_T$ hadron production at midrapidities}
\label{mid}
%
%

\begin{figure}[h]
\resizebox{14.5pc}{!}{
     \includegraphics[height=0.4\textheight]{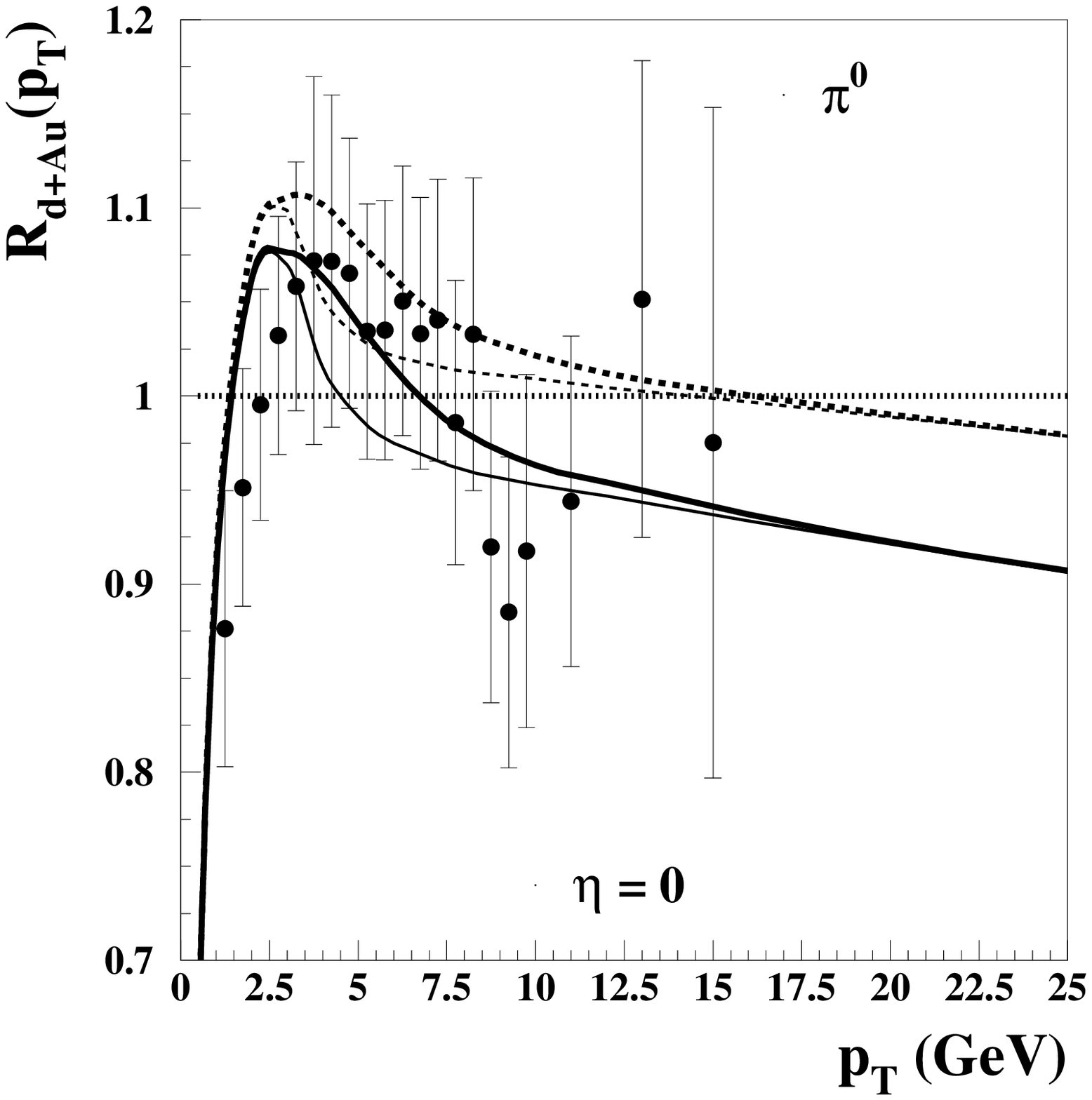}}
\hspace*{0.75cm}
\resizebox{18pc}{!}{
     \includegraphics[height=0.4\textheight]{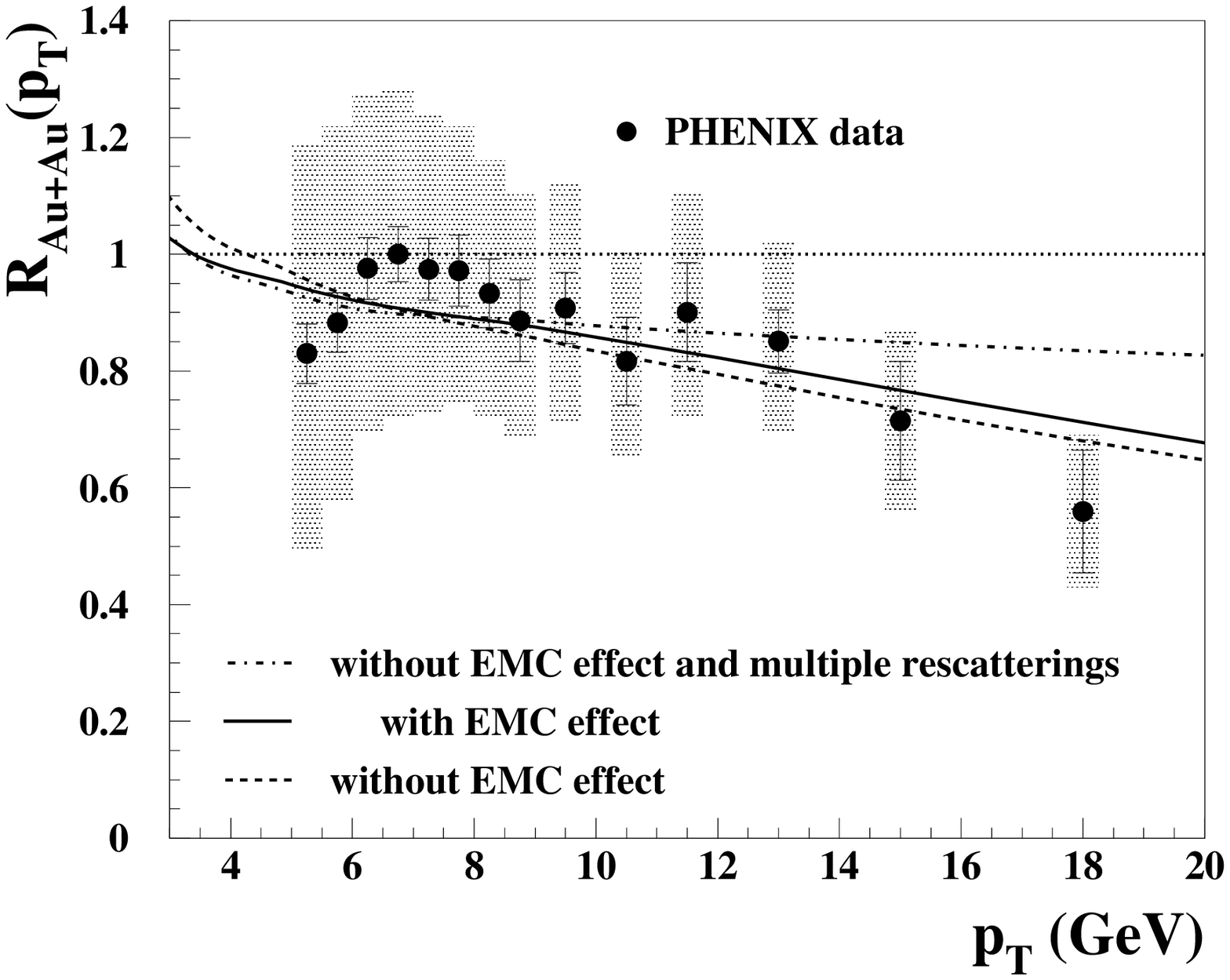}}
\caption
 {
(Left)
Nuclear attenuation factor $R_{d+Au}(p_T)$ as a function of
$p_T$ for production of $\pi^0$ mesons
at $\sqrt{s}=200\GeV$ and $\eta = 0$ vs. data from
PHENIX Collaboration \cite{phenix}. 
(Right)
Nuclear modification factor for direct photon production
in
$Au-Au$ collisions as a function of $p_T$.
}
 \label{phenix}
\end{figure}

As a consequence of $x_F$- scaling is an
expectation of similar nuclear effects also
at midrapidities. 
However, the corresponding values of $p_T$
should be high enough
to keep the same value of $x_F$.
Such an expectation is confirmed by
the recent data from the PHENIX Collaboration \cite{phenix}
showing an evidence for nuclear suppression
at large $p_T > 8\,$GeV
(see the left panel of Fig.~\ref{phenix}). 

At $\eta = 0$ the small-$p_T$ region
is dominated by production and fragmentation of gluons. 
On the other hand, the region of very large $p_T$
is dominated by production and
fragmentation of valence quarks.
Consequently, any value of the hadron
transverse momentum differs only in the relative
contributions of valence quarks and gluons.

It means that we include also gluons 
in our calculations. 
Details can be found in ref.~\cite{knst-01}.
Correspondingly, the cross section for hadron
production, Eq.~(\ref{80}), is extended also
for gluons with corresponding 
distribution function, parton scattering cross section and the 
fragmentation function.
Including multiple parton interactions, the 
gluon distribution in the nucleus is given by the same
formula as for quarks (see Eq.~(\ref{100}),
except $\sigma_{eff}$, which should be higher by the
color factor $9/4$.

If the effects of multiple parton rescatterings are
not taken into account the $p_T$ dependence of 
$R_{d+Au}(p_T)$ is described by the 
thin dashed line. 
One can see from 
the left panel of Fig.~\ref{phenix}
that our calculations at moderate $p_T$ are not in a bad
agreement with data and a small suppression at large $p_T$
is given by the isospin effects. 
After inclusion of multiple
parton rescatterings the model predictions
presented by the thin solid line
underestimate the data 
at moderate $p_T$.
However, at larger $p_T$ quite a strong onset
of nuclear effects is not
in disagreement with corresponding experimental points.
\vspace*{0.2cm}

Calculations in the RHIC energy range at 
midrapidities are most complicated since this is
the transition region between the regimes of long 
(small $p_T$) and short (large $p_T$) coherence lengths.
Instead of too complicated rigorous
light-cone Green function formalism \cite{knst-02,krt2,n-07,n-08}
we preset corrections for finite coherence length using the
linear interpolation performed by means of the
so-called nuclear longitudinal form factor~\cite{knst-01}.
Such a situation is described by the thick solid and
dashed lines reflecting the cases with and without inclusion
of the multiple parton rescatterings, respectively.
It brings the model predictions to a better agreement with data
at moderate $p_T$. Nuclear suppression at large $p_T > 10\,$GeV
observed by the PHENIX experiment \cite{phenix}
can not be explained as a result of CGC
because data cover rather large $x_2\sim 0.05-0.1$. 

%
%
\section{Direct photon production in Au-Au collisions}
\label{dp}
%
%

Expressions for the production cross sections
have been derived employing the dipole formalism
~\cite{kst1,krt1,krtj-03,jkt-01,prepar}.
Model predictions for
$R_{Au-Au}$ as a function of $p_T$ are compared with the
PHENIX data~\cite{phenix-dp} in the right panel of Fig.~\ref{phenix}.
If multiple parton rescatterings are not taken into account
the model calculations depicted by the dash-dotted line
overestimate the data at large $p_T\gsim 13\,$GeV.
The onset of isospin effects gives a value
$R_{Au-Au}\rightarrow 0.8$ in accord with our calculations.
Inclusion of the multiple parton rescatterings leads to a stronger
nuclear effects at large $p_T$ as is demonstrated 
by the dashed line. It brings
a better agreement of the model with data.
Finally, the solid line additionally includes also a
small correction for the EMC effect~\cite{eks-98}.

%
%
\section{Drell-Yan production at large $x_F$}
\label{dys}

\begin{figure}[h]
\resizebox{32pc}{!}{
     \includegraphics[height=0.4\textheight]{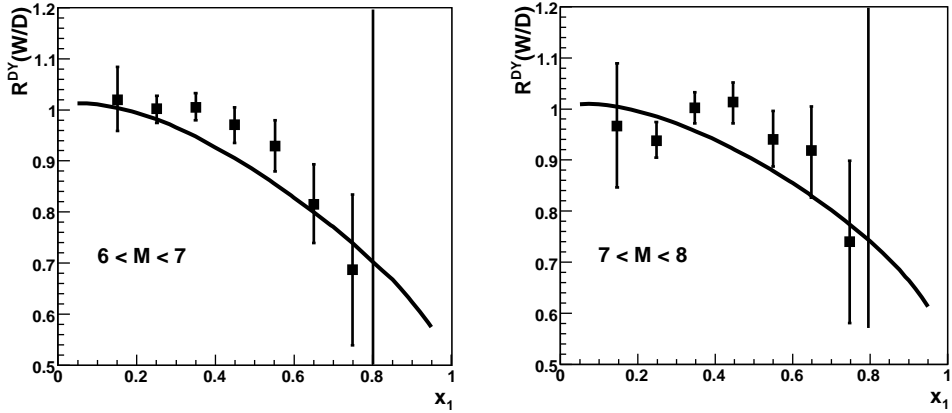}}
\caption
 {
Ratio of Drell-Yan cross sections on Tungsten and
Deuterium as a function of $x_1$. 
}
 \label{dy}
\end{figure}

The DY reaction is also known to be considerably suppressed
at large $x_F$ \cite{johnson} as one can see from Fig.~\ref{dy}.
Model calculations have been performed 
using expressions for the production cross sections
in the color dipole approach
\cite{krt1,krtj-03}.
We included also the effect of multiple parton rescatterings
\cite{knpsj-05,knpsj-05c,npps-08}
discussed above. Model
predictions are in a reasonable agreement with
data from the E772 experiment~\cite{e772}.
\vspace*{-0.1cm}
%
%
\vspace*{-0.2cm}

%
%
\section{Summary and conclusions}\label{conclusions}
%
%

In this paper we analyze a significant nuclear suppression
at forward rapidities (large $x_F$) for several processes.
The new results are the following :

\begin{itemize}

\item
QCD factorization fails at the kinematic limits, $x_F \rightarrow 1$,
$x_1\rightarrow 1$.
Nuclear targets cause a suppression of partons with $x\rightarrow 1$,
due to energy sharing problems.

\item
Suppression of high-$p_T$ hadrons at large rapidity observed by
the BRAHMS and STAR Collaborations is well explained.

\item
We predict $x_1$ ($x_F$) scaling, i.e. the same nuclear effects
at different energies and rapidities corresponding to the
same value of $x_1$ ($x_F$).

\item
Model predictions are in a good agreement
with NA49 data \cite{na49}
and clearly demonstrate
the rise of nuclear suppression with $x_F$.

\item 
Predicted strong nuclear suppression for the large-$p_T$
direct photon production in $Au-Au$ collisions
is in a good agreement with the PHENIX 
data~\cite{phenix-dp}.

\item
According to $x_F$ scaling we predict nuclear suppression
at large $p_T$ also for hadron production at $\eta = 0$.
Model calculations describe well the PHENIX data \cite{phenix}.

\item
Study of nuclear effects
at midrapidities is very important because
at large $p_T$
the data cover rather large $x_2\sim 0.05-0.1$,
where no effect of coherence is possible. 
It allows to exclude the saturation models or
the models based on CGC.
 
\item
Suppression of Drell-Yan pairs at large $x_F$
observed by E772 Collaboration
\cite{e772} is well explained.
\vspace*{-0.1cm}
\end{itemize}


\begin{theacknowledgments}

This work was supported in part by Fondecyt (Chile) grant 1050519,
by DFG (Germany)  grant PI182/3-1, by the Slovak Funding
Agency, Grant No. 2/7058/27 and by the
grant VZ MSM 6840770039, and LC 07048 (Czech Republic).

\end{theacknowledgments}

\end{document}